\documentclass[letterpaper, 10 pt, conference]{ieeeconf}
\IEEEoverridecommandlockouts  
\let\labelindent\relax
\usepackage{enumitem}

\usepackage{mathtools}
\usepackage{url}
\usepackage{hyperref}
\usepackage{graphics} 
\usepackage{epsfig}
\usepackage{cite}
\usepackage{times} 
\usepackage{amsmath,amssymb,amsfonts}
\usepackage{algorithm}
\usepackage{algpseudocode}
\usepackage{graphicx}
\usepackage{textcomp}
\usepackage{varioref}
\usepackage{import}
\usepackage{framed,xcolor}
\usepackage{color}
\usepackage{comment}
\usepackage{epstopdf}   
\usepackage{psfrag}
\usepackage{breqn}
\usepackage{float}
\usepackage{booktabs}
\usepackage{soul}
\usepackage{amsbsy}
\usepackage[bottom]{footmisc}
\usepackage{lineno}
\usepackage{cleveref}
\usepackage{microtype}
\usepackage{comment}
\usepackage{arydshln}
\usepackage{mathtools}
\usepackage{optidef}
\usepackage{graphicx}

\usepackage{tikz}
\usepackage{standalone} % Necessary to skip headers of {standalone} documents
\usepackage{tikzscale} % To use \includegraphics with TikZ code

\usetikzlibrary{backgrounds,fit,positioning}
\usetikzlibrary{shapes.geometric, arrows}
\tikzstyle{arrow} = [thick,->,>=stealth]
\newcommand{\vect}[1]{\ensuremath{\boldsymbol{\mathrm{#1}}}}

\definecolor{dgreen}{rgb}{0.0, 0.5, 0.0}
\definecolor{dpurple}{rgb}{0.5, 0, 0.6}
\definecolor{dorange}{rgb}{0.9, 0.4, 0}

\title{\LARGE \bf Robust Model Predictive Control for   Aircraft Intent-Aware Collision Avoidance}
\author{Arash Bahari Kordabad, Andrea Da Col, Arabinda Ghosh, Sybert Stroeve, and Sadegh Soudjani
\thanks{Arash Bahari Kordabad, Andrea Da Col, Arabinda Ghosh, and Sadegh Soudjani are with the Max Planck Institute for Software Systems, Kaiserslautern, Germany. E-mail: {\tt\small\{arashbk, adacol, arabinda, sadegh\}@mpi-sws.org.} Sybert Stroeve is with the Royal Netherlands Aerospace Centre NLR. E-mail: {\tt\small Sybert.Stroeve@nlr.nl}. \newline This research is supported by the following grants: EIC 101070802 and ERC 101089047.}}
\begin{document}
\bstctlcite{IEEEexample:BSTcontrol}
\maketitle
\thispagestyle{empty}
\pagestyle{empty}
\begin{abstract} This paper presents the use of robust model predictive control for the design of an intent-aware collision avoidance system for multi-agent aircraft engaged in horizontal maneuvering scenarios. We assume that information from other agents is accessible in the form of waypoints or destinations. Consequently, we consider that other agents follow their optimal Dubin's path--a trajectory that connects their current state to their intended state--while accounting for potential uncertainties. We propose using scenario tree model predictive control as a robust approach that demonstrates computational efficiency. We demonstrate that the proposed method can easily integrate intent information and offer a robust scheme that handles different uncertainties. The method is illustrated through simulation results. 
\end{abstract}
%\AG{@Arash, Could you please add legends directly to the figure to make it self-explanatory?}ok, will do.
\section{Introduction}
As air traffic continues to grow, the need for advanced, intent-aware collision avoidance systems for unmanned and manned aircraft becomes increasingly critical. These air-traffic systems determine optimal routes and maneuvers to avoid collisions, using centralized or decentralized strategies~\cite{chen2009hierarchical}. Some key challenges including managing dynamic environments, ensuring real-time response, and handling uncertainties in aircraft behavior, are often addressed by mathematical optimization~\cite{jaeger2013aircraft}. These issues are crucial for advanced air traffic management and the safe integration of diverse aircraft types~\cite{gopalakrishnan2021control}. Research efforts have been dedicated to the development of advanced technologies and protocols to mitigate collision risks and improve the safety of airspace operations~\cite{julian2016policy,zhang2018safety,yasin2020unmanned}.

In order to ensure safe and efficient operations for multi-agent aircraft, having a reliable collision avoidance system (CAS) is crucial. CAS are designed based on centralized or decentralized strategies. Centralized strategies mostly consists of non-cooperative UAVs, manage collision avoidance with a single control entity, and ensures the optimal airspace use in the expense of significant computational resources. On the contrary, decentralized systems offer a cooperative approach which leads to scalability and robustness. Here, UAVs are allowed to independently determine its path while communicating with others. Thus, compared to noncooperative approaches, cooperative methods enhance situational awareness and decision-making~\cite{gao2020unmanned}. Existing literature featured both cooperative and non-cooperative approaches e.g., Palmer et al. proposed both centralized and decoupled approaches for cooperative collision avoidance of UAVs~\cite{palmer2020co}.

%Having a reliable collision avoidance system (CAS) for multi-agent aircraft—where multiple aircraft operate within a shared airspace—is crucial for ensuring safe and efficient operations, especially in crowded airspace. 

In the context of multi-agent control, integration of intent awareness is a cooperative strategy. It involves obtaining and understanding the information of the intentions of humans and/or autonomous agents. This understanding is then integrated into planning and control algorithms~\cite{kelley2010understanding}. Such insight into intent helps in planning and decision-making, allowing for better anticipation and adaptation to the actions of other agents. Many previous methods lack the capability to comprehend the intentions of humans or other agents with whom they interact, potentially resulting in inefficiencies, misinterpretations, or even unsafe behavior in dynamic and uncertain environments~\cite{bowman2019intent}. In the current literature, intent awareness is primarily used in human-robot interaction, where the main goal for an autonomous agent is to comprehend human intentions and execute designated tasks~\cite{ranatunga2015intent, huang2022intent}. However, in the context of multi-agent planning and control, especially in air traffic management of UAVs, the use of intent awareness is not explored, which serves as the primary motivation for this paper.

On the industrial side, airborne collision avoidance systems (ACAS) are critical safety measures designed to prevent mid-air collisions between aircraft. These systems use transponder signals to detect the presence of other aircraft, calculate potential collision threats, and provide pilots with real-time avoidance instructions. The ACAS X program represents the next generation of these systems, introducing enhanced algorithms and more advanced technology to improve accuracy and reduce unnecessary alerts. ACAS X includes various versions tailored for different aviation sectors, such as ACAS Xa for large aircraft and ACAS Xu for unmanned aerial systems~\cite{kochenderfer2012next, kochenderfer2011robust}. The threat resolution module in ACAS X employs optimized decision logic tables that are generated offline using Dynamic Programming (DP)~\cite{stroeve2023matters}. DP optimizes decision-making under uncertainty, breaking down complex problems into manageable subproblems to find the optimal solution efficiently. It plays a crucial role in environments where the outcomes are uncertain and when interacting with other agents whose behaviors are dynamic~\cite{bertsekas2012dynamic}. However, it might be computationally complex for large-scale systems~\cite{powell2007approximate}. Other stochastic model-based methods such as Markov Decision Process framework, have also been applied to UAV systems in both single-agent~\cite{trotti2024towards} and multi-agent~\cite{trotti2024markov} scenarios.

Model Predictive Control (MPC) offers the advantage of real-time optimization by solving control problems at each time step, allowing for dynamic adjustment based on current state information. Unlike DP, MPC can handle constraints and adapt to changes in the system model and environment efficiently. This makes MPC more suitable for complex, real-time applications. Scenario-tree MPC is a robust MPC technique designed to handle finite and discrete uncertainties in a nonlinear system~\cite{klintberg2016improved}. It extends traditional MPC by explicitly considering multiple possible future scenarios, typically represented in a tree structure. Scenario-tree MPC offers a computationally tractable strategy for handling uncertainty in future discrete scenarios, allowing for approximate robust decision-making~\cite{kordabad2021reinforcement}. Scenario-tree MPC can then potentially provide an approximate solution for the DP~\cite{kordabad2023equivalence}.

%For multi-agent systems, An intent-aware optimal collision avoidance and trajectory planning has been proposed by the authors in~\cite{biswas2022intent} for a pursuit vehicle. In the context of UAVs, a neural network model has been introduced in~\cite{li2024behavior} to achieve UAV intent recognition by converting long-term temporal behavioral data into natural language. 

%In the context of multi-agent control, intent awareness is a cooperative approach that refers to understanding or having the information of the intentions of other agents, including humans or autonomous agents, and incorporating this understanding into planning and control algorithms~\cite{kelley2010understanding}.

%In~\cite{huang2022intent}, an intent-aware interactive ambient intelligence system was proposed that enables robots to infer human intents actively and assists humans in accomplishing daily tasks. A few strategies for designing intent-aware multi-agent systems to enhance the implicit coordination of human-agent teams were provided in \cite{schneider2020operationalized}.  

%Intent awareness finds its primary application in the realm of human-robot interaction. A novel adaptive admission controller was proposed in~\cite{ranatunga2015intent} to integrate human intent, enhancing the efficacy of the physical human-robot interaction. A haptic shared control paradigm was presented in~\cite{ly2021intent} to modulate the level of robotic guidance based on predictions of human motion intentions. 

In this paper, we consider the horizontal maneuvering of an ownship-intruder aircraft system. Our primary objective is to incorporate intent information into the planning and control of collision avoidance systems for the system in shared airspace. We first formulate intent awareness as information on waypoints or destinations available from other agents. From ownship perspective, the intruder is assumed to follow an optimal Dubins path—i.e., a shortest-length trajectory subject to curvature constraints—connecting its current state to its intended waypoint. A robust MPC strategy is then {proposed that directly integrates intent information using a scenario-tree MPC approach. Specifically, we model uncertainties in the intruder’s angular velocity within a branching scenario tree, thereby generating a finite set of possible future evolutions of the system. By leveraging this representation, our method approximates the robust optimal policy while maintaining computational tractability.

The paper is structured as follows. Section~\ref{sec:back} presents a background on the classic MPC and scenario-tree MPC  approaches. Section~\ref{sec:horiz:CAS} describes the equations of motion and dynamics for horizontal maneuvering UAVs. Section~\ref{sec:intent} elaborates on intent awareness and demonstrates how it can be integrated into the CAS. Moreover, an intent-aware robust MPC is presented in this section. Section~\ref{sec:simulation} provides the simulation results, and section~\ref{sec:con} delivers a conclusion.

\section{Background}\label{sec:back}
In this section, we provide a background on the classic and scenario tree MPC approaches. 

\subsection{Model Predictive Control (MPC)}
Model Predictive Control (MPC) is a powerful optimization-based control method widely used in both academic research and industrial applications~\cite{cai2021mpc}. At its core, MPC computes control actions by solving an optimal control problem at each discrete time step, observing the current system state $\vect s$. The optimization is performed over a finite prediction horizon, and only the first optimal input in the resulting control sequence is implemented. This process is repeated at every time step, which gives MPC its characteristic receding horizon or rolling horizon nature~\cite{rawlings2017model}.

One of the main advantages of MPC lies in its ability to handle state and input constraints directly within the optimization problem. For deterministic systems, MPC can often be implemented efficiently in real time, making it suitable for a wide range of practical applications. In its standard form for deterministic dynamics, the MPC problem is formulated as:
\begin{subequations}
\label{eq:MPC0} 
\begin{align}
\min_{\hat{\vect s},\hat{\vect a}}&\quad V_{\text{f}} \left(\hat{\vect s}_N\right) + \sum_{k=0}^{N-1}\, L\left(\hat{\vect s}_k,\hat{\vect a}_k\right) \label{eq:cost0}\\
\mathrm{s.t.} &\quad \forall k\in\{0,\ldots, N-1\}:\nonumber \\&\quad \hat{\vect s}_{k+1} = \vect f_d\left(\hat{\vect s}_k,\hat{\vect a}_k\right),\quad \hat{\vect s}_0 = \vect s \label{eq:dyn0}\\
&\quad \vect h\left(\hat{\vect s}_k,\hat{\vect a}_k\right) \leq 0,\quad \hat{\vect a}_k \in \mathcal{A}, \quad  \label{eq:const0} %  \label{eq:IC0}
\end{align}
\end{subequations}
where $L$ is the stage cost, $V_{\text{f}}$ is the terminal cost, $\vect f_d$ is the deterministic dynamics model, $\vect h$ is the constraints, $N$ is the horizon length and $\mathcal{A}$ is the control set. In order to distinguish between the actual system trajectory and the predicted state-input profile, we use the notation \textit{hat} $\hat \cdot$ for the latter. The open-loop optimization~\eqref{eq:MPC0} is solved recursively at each state $\vect s$ and  produces a complete profile of control inputs $\hat{\vect a}^\star = \{\hat{\vect a}_0^\star,\ldots, \hat{\vect a}_{N-1}^\star\}$ and corresponding state predictions $\hat{\vect s}^\star= \{\hat{\vect s}_0^\star,\ldots, \hat{\vect s}_{N}^\star\}$. The notation \textit{star} $\cdot ^\star$ is used to refer to the optimal value of the decision variables. To incorporate feedback, only the first element $\hat{\vect a}_0^\star$ of the input sequence $\hat{\vect a}^\star$ is applied to the system, a successor state $\vect s^+$ is attained in the next time instance, and the optimization \eqref{eq:MPC0} is solved again for the new state $\vect s^+$. The resulting control policy is thus given by:
\begin{equation*}
%\label{eq:MPC0:Policy} 
\vect\pi_{\mathrm{MPC}}\left(\vect s\right) = \hat{\vect a}_0^\star.
\end{equation*} 
Despite its strengths, this standard deterministic MPC formulation is limited in its ability to handle uncertainties and stochastic effects. Therefore, it is limited in situations where there are no significant uncertainties and relatively small disturbances in the system, as it typically relies on deterministic models. Consequently, it is necessary to employ a notion of robust MPC for nonlinear systems that is computationally efficient. Note that, based on the online receding horizon nature of the MPC, there are circumstances where traditional MPC can still perform acceptably, even in stochastic systems. This is often the case when the system uncertainties are relatively small, predictable, or can be well approximated by the deterministic model used by the MPC. Nevertheless, to ensure reliable performance in the presence of uncertainty—particularly in safety-critical applications—it becomes essential to extend MPC to incorporate robustness explicitly. In the following section, we introduce an efficient and robust MPC approach that addresses these challenges while preserving computational tractability.

\subsection{Scenario-tree MPC}\label{sec:RMPC}
In this section, we present scenario-tree MPC as a tractable and effective method to handle uncertainties in nonlinear systems. Scenario-tree MPC offers a principled way to incorporate multiple possible realizations of system uncertainty while maintaining computational feasibility.
 
Consider a stochastic system where the dynamics depend on an uncertain disturbance $\vect w$ drawn from a known distribution $W$. To approximate the effect of this uncertainty, we sample $m$ disturbance realizations $\vect w^i$ from $W$, with $i \in \{1, \ldots, m\}$. The system dynamics under each sampled disturbance can then be expressed as: 
\begin{equation}
  \vect s^+= \vect f(\vect{s}, \vect{a},\vect w^i),\quad i\in\{1,\ldots, m\}.   
\end{equation}
At each decision point in time, this leads to a branching of the future state trajectory into multiple scenarios, one for each possible disturbance realization. The number of trajectories (scenarios) grows exponentially with the length of the MPC horizon. 

To address this, scenario-tree MPC introduces the concept of a robust horizon, denoted by $N_r < N$. This is the point beyond which the uncertainty is no longer considered explicitly in the scenario branching. Instead, after $N_r$ time steps, the uncertainty is fixed to a nominal or representative value (e.g., $\vect w^1$). This truncation of the scenario tree significantly reduces the exponential growth in the number of branches, ensuring computational tractability while still capturing the most critical early-stage uncertainty in the decision-making process.

Figure~\ref{fig:rmpc} illustrates the evolution of the system represented as a scenario-tree for $m=3$, $N_r=2$, and $N=3$. Each scenario corresponds to a specific trajectory, starting from the root (current state) to a leaf. Each node in the tree represents a possible state of the system at a future time step, and each branch represents a possible transition due to the uncertain parameter $\vect w^i$ with $i\in\{1,2,3\}$. The uncertain parameter is fixed to $\vect w^1$ after $N_r=2$ steps. 

\begin{figure}[ht!]
\centering
\includegraphics[width=0.48\textwidth]{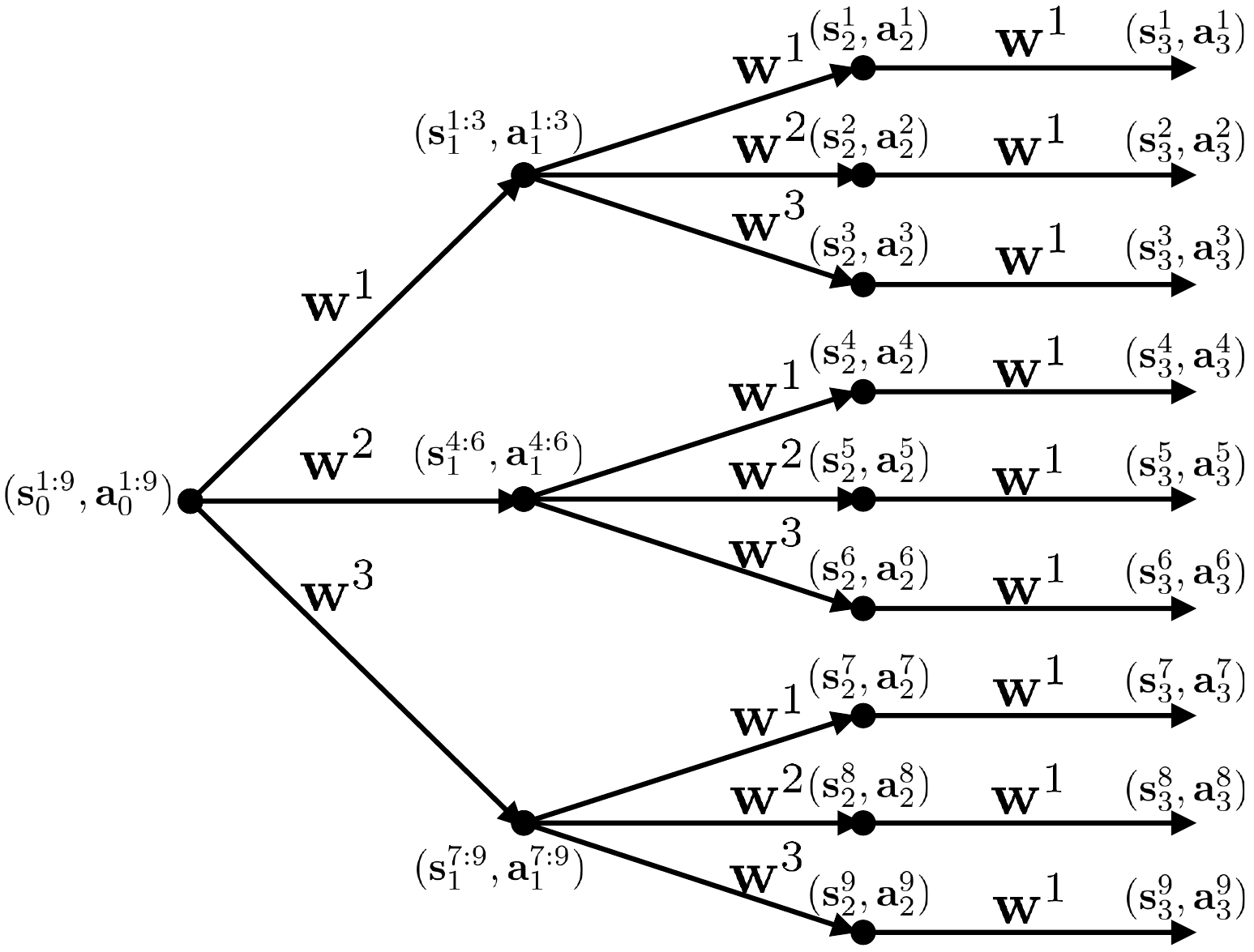}
\caption{The evolution of the system represented as a scenario-tree.}
\label{fig:rmpc}
\end{figure}

The number of all scenarios is $M=m^{N_r}$, and in the figure, there are $9$ different scenarios. The superscripts in the states denote a specific scenario, while the subscripts are used for the time index. The notation $\vect s^{i:j}_k$ denotes $\vect s^{i}_k=\vect s^{i+1}_k=\ldots=\vect s^{j}_k$. For instance, in the beginning, all scenarios start with the current specific state that is denoted by $\vect s_0^{1:9}$.
Furthermore, since uncertainty cannot be predicted in advance, control actions must depend solely on historical realizations of uncertainty. Hence, $\vect{a}^j_{i} = \vect{a}^l_{i}, \ \forall i = 0, \ldots, k $, if the uncertainty realization for scenarios $j$ and $l$ are identical up to and including time stage $k$. This restriction is commonly referred to as a \textit{non-anticipativity} constraint~\cite{klintberg2016improved}. This constraint in Figure~\ref{fig:rmpc} can be  expressed as $\vect a^{1:9}_0$, $\vect a^{1:3}_1$, $\vect a^{4:6}_1$, and $\vect a^{7:9}_1$. 

Considering all scenarios in the constraints and optimizing the cost function based on their average, the scenario-tree MPC optimization can be expressed as follows:

%\begin{subequations}
%\label{eq:MPC1} 
\begin{align*}
\min_{\hat{\vect s},\hat{\vect a}}&\quad \sum_{j=1}^M\left(V_{\text{f}} \left(\hat{\vect s}^j_N\right) + \sum_{k=0}^{N-1}\, L\left(\hat{\vect s}^j_k,\hat{\vect a}^j_k\right)\right), \\
\mathrm{s.t.}   &\quad \forall k\in\{0,\ldots, N-1\},\, \forall j\in\{1,\ldots, M\}:\nonumber  \\&\quad \hat{\vect s}^j_{k+1} = \vect f\left(\hat{\vect s}^j_k,\hat{\vect a}^j_k, \vect w^i\right),\quad \hat{\vect s}^j_0 = \vect s \\
&\quad \vect h\left(\hat{\vect s}^j_k,\hat{\vect a}^j_k\right) \leq 0,\quad \hat{\vect a}^j_k \in \mathcal{A}, \quad 
 \\
&\quad \forall l\in\{1,\ldots, M\},  \, \forall i\in\{0, \ldots, k\}: \nonumber
\\&\quad  \hat {\vect{a}}^j_{k} = \hat{\vect{a}}^l_{k}\quad \text{if}\quad \hat {\vect{s}}^j_{i} = \hat{\vect{s}}^l_{i}, 
  %  \label{eq:IC0}
\end{align*}
%\end{subequations}
where
\begin{equation*}
    i=\left\{\begin{matrix}
{\mathrm{mod}(\left \lceil \frac{j}{m^{N_r-k-1}} \right \rceil,m)} &  k< N_r  \\
1 & N_r\leq k,  \\
\end{matrix}\right.
\end{equation*}
indicates the index of parameter $\vect w^i$ for each time-scenario pair, and where the function $1\leq \mathrm{mod}(n,m)\leq n$ is the remainder of $n/m$ if $n/m$ is not an integer, otherwise $\mathrm{mod}(n,m)=m$. Moreover, $\left \lceil n \right \rceil:=\min \{m\in \mathbb{Z} \, |\, m\geq n\}$.

Analogous to the standard MPC, the policy of robust MPC based on a scenario-tree is determined as follows:
\begin{equation*}
\vect\pi_{\mathrm{RMPC}}\left(\vect s\right) = \hat{\vect a}_0^{1:M,\star},
\end{equation*} 
where $ \hat{\vect a}_0^{1:M,\star}$ is the optimal solution of $\hat{\vect a}_0^{1}=\ldots=\hat{\vect a}_0^{M}$.
\section{Equations of Motion}\label{sec:horiz:CAS}
In this section, we provide the equations of motion for two aircraft maneuvering in the horizontal plane. We consider two aircraft, ownship and intruder, to have the following equation of motion in the discrete-time setting:
\begin{equation}\label{eq:single agent EoMs1}
    \vect s_{k+1}^i:=\begin{bmatrix}
        {x}^{i}_{k+1} \\ {y}^{i}_{k+1} \\ {\sigma}^{i}_{k+1}
    \end{bmatrix} = \begin{bmatrix}
        {x}^{i}_{k} \\ {y}^{i}_{k} \\ {\sigma}^{i}_{k}
    \end{bmatrix}+t_e
    \begin{bmatrix}
        v^{i}_k \cos{\sigma^{i}_{k}} \\ v^{i}_k \sin{\sigma^{i}_{k}} \\ u^{i}_k 
    \end{bmatrix},\quad i\in\{1,2\},
\end{equation}
where the superscript $i=1$ stands for the ownship whereas $i=2$ corresponds to the intruder, $\vect s_k^i$ is the state of the aircraft $i$ at time $k$, ${x}^{i}_{k}$ and ${y}^{i}_{k}$ are the position states, and ${\sigma}^{i}_{k}$ is the heading angle. The constant $t_e$ is the sampling time and is assumed to be $1$ sec, and $v^{i}_k$ and $u^{i}_k$ are the linear and angular velocities, respectively. Figure~\ref{fig:1} shows the geometry of the problem.

%In horizontal maneuvering with an ownship-intruder setting, it is preferable to describe the system based on relative dynamics, especially when using methods such as dynamic programming, which have high computational complexity. In this approach, the state of the system is defined as~\cite{Julian}:
%\begin{equation}\label{eq:state}
 %   \vect{s} =  \begin{bmatrix}     \rho & \theta & \psi & v^1 & v^{2} & \tau & s_{\mathrm{adv}} \end{bmatrix}^\top\,,
%\end{equation}
%where $\rho$ is the distance between aircraft, $\theta$ is the bearing angle to intruder, $\psi$ is relative heading of intruder, $\tau$ is the time to loss of the vertical separation and $s_{\mathrm{adv}}\in \mathcal{A}$ is the advisory taken in the previous time step. 
\begin{figure}[ht!]
\centering
\includegraphics[width=0.45\textwidth]{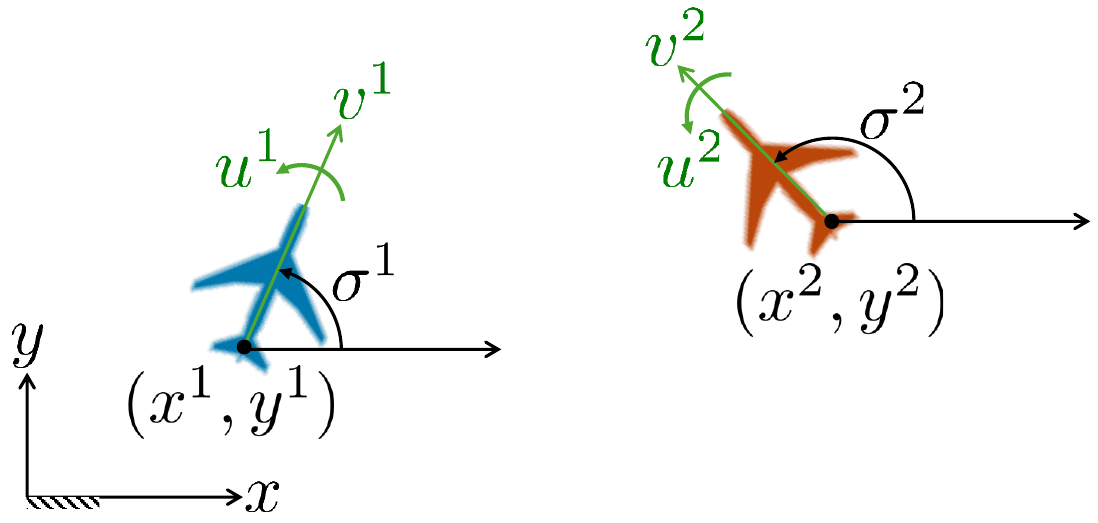}\caption{Geometry of two aircraft in the horizontal plane in the earth-fixed coordinate system. The black variables are the state variables and the greens are the velocities.}
\label{fig:1}
\end{figure}
The 2D model in~\eqref{eq:single agent EoMs1} focuses on the kinematics of horizontal motion, which is standard in high-level collision avoidance, as used for example in~\cite{panoutsakopoulos2022towards}. For the scope of this paper, it offers a clear and effective way to study how intent information can be incorporated into planning and control without introducing unnecessary model-related complexity.

\textbf{Action space ($\mathcal{A}$):} 
%In the horizontal plane, the action set includes turning right, turning left, and Clear-of-Conflict (COC). The advisories are outlined in Table~\ref{table:action5:h}.
%\begin{center}
%\begin{table}[ht]
%\caption{Horizontal advisory set~\cite{julian2019guaranteeing}}
%\begin{tabular}{l l l}
% \hline
% Advisory & Description & Ownship Turn Rate  \\ 
% \hline
% COC & Clear-of-conflict  & $[-1.5^\circ/s,\,1.5^\circ/s]$ \\ 
% WL & Weak Left  & $[1.0^\circ/s,\,2.0^\circ/s]$  \\
% WR &Weak Right  & $[-2.0^\circ/s,\,-1.0^\circ/s]$  \\
% SL & Strong Left  & $[2.0^\circ/s,\,4.0^\circ/s]$  \\
% SR & Strong Right & $[-4.0^\circ/s,\,-2.0^\circ/s]$  \\
% \hline
%\end{tabular}
%\label{table:action5:h}
%\end{table}
%\end{center}
We consider the linear and angular velocities as control inputs. Although discrete advisories (i.e., control inputs) are used more frequently in the DP approach~\cite{julian2019guaranteeing}, the MPC approach requires solving an optimization problem at each time instance. In the MPC scheme, it is preferable to have decision variables in the continuous space to avoid the computational complexity associated with discrete variables and mixed-integer programming. Therefore, a continuous set $\mathcal{A}=[6\,,\,9][\mathrm{m}/s]\times [-0.1\,,\,0.1][\mathrm{rad}/s]$ is considered for the ownship, and $\mathcal{A}=10[\mathrm{m}/s]\times[-0.07\,,\,0.07][\mathrm{rad}/s]$ is used for the intruder.

\textbf{Uncertainties:} Uncertainties in aircraft dynamics typically manifest additively in the angular velocity. Specifically, the actual advisory is typically considered to be the nominal value with a relatively high probability (e.g., $0.5$), and the endpoints of the intervals have a relatively lower probability (e.g., $0.25$). This type of uncertainty is used in DP approaches due to its discrete nature, making it easier to address numerically~\cite{julian2019guaranteeing}. In this paper, we consider a similar way of uncertainty modeling for the intruder with equal probability for the scenario-tree MPC approach as a robust MPC. More specifically, we assume that the intruder may adopt either a nominal angular velocity or the angular velocities at the boundaries to form the branches of the scenario-tree MPC.

\section{Intent-Aware Scenario-tree MPC }\label{sec:intent}
In this paper, one of our main contributions is the consideration of scenarios where the ownship is aware of the intruder's intent in the MPC scheme. This intent is modeled as the intruder's future path, or equivalently, its upcoming control inputs, which are available to the ownship. Specifically, the intent is represented as a series of waypoints that the intruder will pass through. Using these waypoints, the overall path is determined based on the Dubins path, which is the optimal path that connects the intruder's current position to the next waypoint~\cite{dubins1957curves}.

The Dubins path connects an initial state (position and direction) to the next waypoint using a sequence of simple motion primitives that respect curvature constraints. Specifically, it is composed of segments that represent basic maneuvers: turning left (L) refers to a circular arc with maximum allowed angular velocity in the counterclockwise direction; moving straight (S) corresponds to linear motion at a constant heading; and turning right (R) denotes a circular arc with maximum angular velocity in the clockwise direction. These segments are combined in sequences such as LSR or RSL to form the shortest feasible path}. Figure~\ref{fig:dubin} illustrates an example of such a path, denoted as LSR. The gray dashed-line circles represent the maximum possible curvature based on the limitations of linear and angular speed. Both geometric~\cite{anisi2003optimal} and analytical methods~\cite{bui1994shortest} are available to compute this optimal path.
   \begin{figure}[ht!]
                \begin{center}
                    \includegraphics[width=0.95\linewidth]{f3.tex}
                \end{center}
            \caption{An LSR Dubins path: An optimal connecting path of the initial state $\vect s_0$ to the target state $\vect s_T$.}
            \label{fig:dubin}
            \end{figure}

More specifically, the optimal Dubin's path provides a mapping from the initial-target state pair to the path connecting these two states, i.e.,  $(\vect s_0,\vect s_T)\rightarrow (\vect s_0,\vect s_1,\ldots, \vect s_T)$ and consequently the angular velocities can be obtained by a certain time-varying mapping, as follows:
\begin{equation*}
    u_k=D(\vect s_0,\vect s_T,k), \quad 0\leq k<T,
\end{equation*}
where $D$ is a mapping from the initial state, the target state, and the time index $k$ to the corresponding optimal angular velocity $u_k$, generating the optimal Dubin's path. This mapping can be obtained based on minimizing the path through different types of possible connecting paths including RSR, RSL, LSR, LSL, RLR, or LRL. Note that, the linear velocity is assumed to take a constant value. A detailed computation of this mapping is provided e.g.,  in~\cite{shkel2001classification}. In \textsc{Matlab}, the \texttt{dubinsConnection} function from the Navigation Toolbox can generate such paths using a function call.
            
In the context of MPC, incorporating the intruder's intent information into the control synthesis for the ownship is relatively straightforward. First, the intruder's current state and waypoint are used to determine its Dubin path, which allows for the prediction of the intruder's upcoming control inputs. An MPC scheme is then formulated for the ownship with objectives such as minimizing control effort or reaching a destination, while adhering to constraints such as maintaining a minimum safe distance from the intruder. Consequently, the intruder's dynamics and predicted control inputs are integrated into the ownship MPC scheme. By utilizing the intruder's intent information, these values can be accurately included in the ownship MPC, reducing uncertainty. In contrast, without knowledge of the intruder's intent, predictions are based on assumptions like a direct path for the intruder, which introduces more uncertainty.

While full knowledge of the intruder’s intent may seem strong, in many practical settings—such as cooperative missions, or systems with efficient communication protocols—such information (e.g., waypoints or destinations) can be shared in advance. We show how this information can be incorporated into the the control synthesis, and demonstrate that leveraging intent enables us to improve planning efficiency and maintain safety under uncertainty. We acknowledge that in more complex scenarios, the intruder’s behavior may be partially known or unknown, requiring intent estimation. However, in this work, we focus on the case where intent is known possibly with uncertainty, and we capture this uncertainty explicitly using a scenario-tree MPC approach.

For the ownship ($i=1$), we consider $v^{i}_k$ and $u^{i}_k$ as the control input. These velocities are assumed to be obtained based on Dubin's optimal path for the intruder ($i=2$). Therefore, based on the initial state and the \textit{intent} state for the intruder, we can obtain the corresponding Dubin's path and the corresponding control inputs. Then, the aim is to steer the ownship to its intent state while avoiding collision with the intruder using an MPC method. 

In order to incorporate uncertainties that arise from the intruder, when the ownship is at the state $\vect s_t$, $0\leq t\leq T$, we utilize an intent-aware scenario-tree MPC approach as follows:
\begin{subequations}
\label{eq:RMPC}
\begin{align}
    &\min_{u_{0:N-1}^1,v_{0:N-1}^1,\vect s_{0:N}^1} \,\,\,\,  (\vect s^1_{N}-\vect s_T)^\top Q_{\mathrm{f}}(\vect s^1_{N}-\vect s_T)\label{eq:RMPC:cost}\\  &\qquad\quad\qquad\quad\,\,\,\,+(\vect s^1_{0}-\vect s_T)^\top Q(s^1_{0}-\vect s_T)
    \nonumber\\
    &\quad\quad\quad+\sum_{k=1}^{N-1}  (\vect s^1_{k}-\vect s_T)^\top Q(s^1_{k}-\vect s_T)+R(u^{1}_k-u^{1}_{k-1})^2,\nonumber\\
   &\qquad \mathrm{s.t.} \,\,  \forall j\in\{1,\ldots, M\},  \forall k\in\{0,\ldots, N-1\}:\nonumber \\  &\qquad\quad\,\,\,\eqref{eq:single agent EoMs1},\qquad \forall i\in\{1,\{2,j\}\}, \label{eq:RMPC:dyn}\\
    &\qquad\quad\,\,\, \rho\leq \sqrt{(\vect s^1_{k}-\vect s^{2,j}_{k})^\top \mathrm{diag}(1,1,0) (\vect s^1_{k}-\vect s^{2,j}_{k})},\label{eq:RMPC:safe}\\
  &\qquad\quad\,\,\,  \underline{u}^{1}\leq u^{1}_k\leq \bar{u}^{1},\quad \underline{v}^{1}\leq v^{1}_k\leq \bar{v}^{1},\label{eq:RMPC:own}\\
  &\qquad\quad\,\,\, \text{if}\,\, k< N_r:\,\,\, u^{2,j}_k= \label{eq:RMPC:intr11}\\ &\qquad\quad\,\,\, \left\{\begin{matrix} \bar{u}^{2} &  \text{if}\, \mathrm{mod}(\left \lceil \frac{j}{3^{N_r-k-1}} \right \rceil,3)=0\, \\ 
\underline{u}^{2} &  \text{if}\, \mathrm{mod}(\left \lceil \frac{j}{3^{N_r-k-1}} \right \rceil,3)=1\, \\
  D(\vect s^2_0,\vect s^2_T,t+k) &  \text{if}\, \mathrm{mod}(\left \lceil \frac{j}{3^{N_r-k-1}} \right \rceil,3)=2,
\end{matrix}\right.\nonumber\\ 
 &\qquad\quad\,\,\, \text{if}\,\, k\geq N_r:\,\,\, u^{2,j}_k=D(\vect s^2_0,\vect s^2_T,t+k),\label{eq:RMPC:intr12} \\ 
  &\qquad\quad\,\,\, v^{2,j}_k=\bar{v}^{2}, \label{eq:RMPC:intr2}
 \\ &\qquad\quad\,\,\, \vect s^1_0=\vect s_t,\label{eq:RMPC:ini}
\end{align}
\end{subequations}
where $Q_{\mathrm{f}}$ and $Q$ are positive definite matrices, $R$ is a positive constant, $\rho$ is the minimum allowed horizontal distance of the two aircraft and $\underline{v}^{i}$ ($\underline{u}^{i}$) is the minimum and $\bar{v}^{i}$ ($\bar{u}^{i}$) is the maximum allowed linear (angular) velocity for $i\in\{1,2\}$. 

Robust MPC~\eqref{eq:RMPC} steers the ownship state to its target state $\vect s_T$ with a quadratic cost in~\eqref{eq:RMPC:cost}  while avoiding collision with the intruder, enforced by the constraint~\eqref{eq:RMPC:safe}, which is taking its optimal Dubin path with additional uncertainties. The last term in the cost function in~\eqref{eq:RMPC:cost} minimizes the variations in the angular velocity. Constraint~\eqref{eq:RMPC:dyn} represents the dynamics of the ownship $(i=1)$, and dynamics of the intruder for $M$ different scenarios $i\in\{2,j\},\, j\in\{1,\ldots, M\}$ conducted according to Figure~\ref{fig:rmpc}. Constraint~\eqref{eq:RMPC:own} enforces the control limitations for the ownship. Constraint~\eqref{eq:RMPC:intr11} generates different inputs (as uncertainties) for three branches ($m=3$) of the scenario-tree MPC. We have considered an uncertainty for the angular velocity of the intruder in the form of $u_k^{2}\in\{\underline{u}^{2}, D(\vect s^2_0,\vect s^2_T,k), \bar{u}^{2}\}$ until the robust horizon $N_r$ with $\underline{u}^{2}=-0.07~[\mathrm{rad}/s]$ and $\bar{u}^{2}=0.07~[\mathrm{rad}/s]$. After the robust horizon $N_r$, the intruder follows its optimal dubin's path and \eqref{eq:RMPC:intr12} enforces this. The linear velocity for the intruder is assumed to be constant and follows its maximum value, as represented in constraint~\eqref{eq:RMPC:intr2}. Finally, constraint~\eqref{eq:RMPC:ini} sets the initial state of the ownship MPC to the actual current state $\vect s_t$ and the MPC is solved in the receding horizon manner. Note that in MPC~\eqref{eq:RMPC}, because the uncertainties appear in the control input value, there is no need to add the non-anticipativity constraint. This restriction is implicitly accounted for in constraint~\eqref{eq:RMPC:intr11}.

In other words, these $M$ scenarios account for uncertainties in the intruder's execution of angular velocity (i.e., control input) by considering both the nominal angular velocities that follow the optimal Dubins path and the worst-case angular velocities over a few steps ahead (robust horizon). Incorporating these scenarios into the optimization~\ref{eq:RMPC} reduces the feasibility domain, leading to more conservative control inputs for the ownship compared to when only the intruder's nominal trajectory is considered. However, this approach enhances robustness against potential uncertainties arising from e.g.,  human error, mismatches in the intent information, external disturbances, etc.

In cases where intent information is not available or accounted for, a straight-line prediction is typically assumed for the intruder. Thus, in MPC~\eqref{eq:RMPC}, the value of the function $D$ is replaced with zero. This discrepancy between the prediction and the actual path of the intruder can result in a non-optimal path, as it will be demonstrated in the simulation results. It is important to note that considering intent information is beneficial not only for optimization and safety purposes but also for the MPC recursive feasibility. Recursive feasibility in MPC schemes can be challenging when the system is uncertain, such as when the intruder's waypoint does not align with the predicted path. Knowledge of the intruder's intent assists in reducing the uncertainty and, consequently, achieving recursive MPC feasibility.

Note that, while this paper primarily focuses on the two-aircraft ownship-intruder scenario, the approach can be extended naturally to cases with more than two aircraft. For multi-aircraft scenarios, the computational complexity is affected only by the number of constraints, which grows linearly with the number of aircraft. In other words, adding more aircraft only increases the number of constraints, without introducing additional sources of complexity.

\section{Simulation results}\label{sec:simulation}
In the next section, we illustrate simulation results for the collision avoidance problem for two aircraft based on the proposed method. Throughout the simulations, we have used a $30$s horizon length for the MPC ($N=30$) and a robust horizon of $N_r=3$.

\begin{figure}[ht!]
\centering
\includegraphics[width=0.48\textwidth]{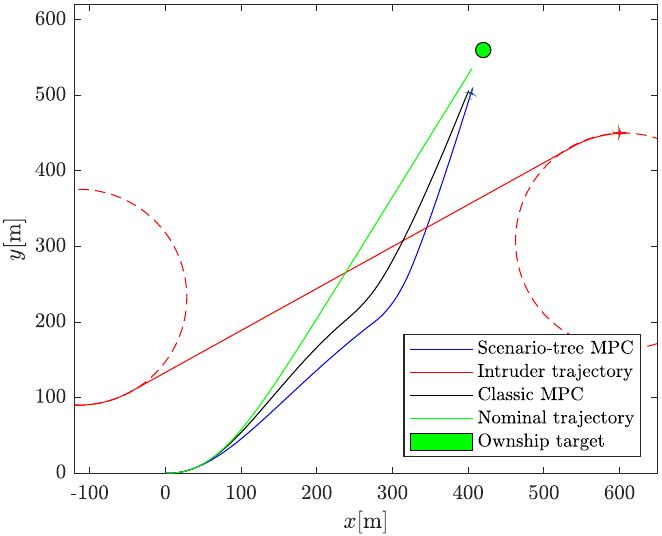}
\caption{Comparing the classic MPC trajectory (black), the scenario-tree MPC trajectory (blue), and the nominal trajectory (green) without safety constraints for the ownship. The intruder Dubins path is shown in red and the green area is the destination of the ownship.}
\label{fig:MPC:0}
\end{figure}

Figure~\ref{fig:MPC:0} compares the ownship nominal trajectory, the path obtained from the classic MPC, and the scenario-tree MPC. In the nominal trajectory, without any intruder present, the ownship follows an LS Dubin optimal path to reach its next waypoint. However, when an intruder exists in the airspace, the path deviates from the optimal nominal path to avoid collisions. The intruder and its corresponding Dubins path are shown in red.  It can be seen that both the scenario-tree MPC and the classic MPC successfully avoid collisions with the nominal trajectory of the intruder. However, the scenario-tree approach maintains a greater distance from the nominal path of the intruder compared to the classic MPC approach.

\begin{figure}[ht!]
\centering
\includegraphics[width=0.48\textwidth]{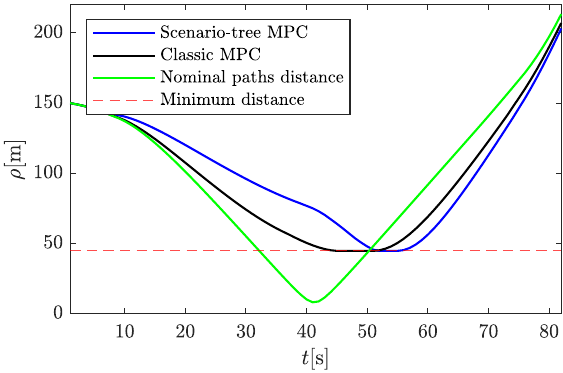}
\caption{The distance between the ownship and the intruder over time. The red dashed line is the minimum allowed distance. The green curve is for the nominal path.}
\label{fig:MPC:2}
\end{figure}

\begin{figure}[ht!]
\centering
\includegraphics[width=0.48\textwidth]{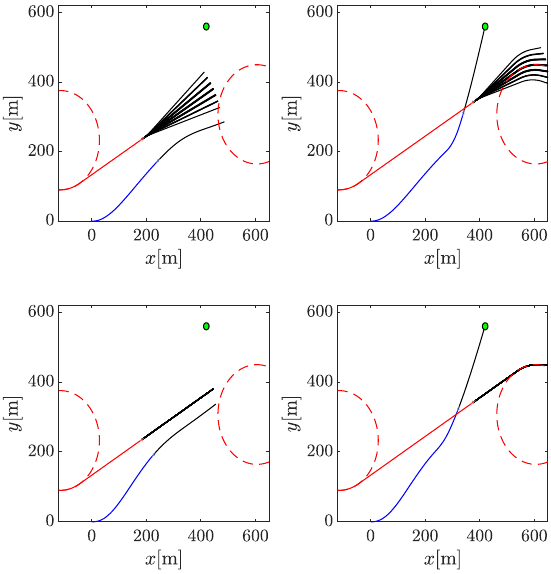}
\caption{The top-left (bottom-left) figure shows the trajectories at $t=35$s (before passing behind the intruder), and the top-right (bottom-right) figure shows the trajectories at $t=57$s, after passing behind the intruder, for the scenario-tree (classic) MPC approach.}
\label{fig:MPC:3}
\end{figure}

Figure~\ref{fig:MPC:2} shows the distance between the ownship and the intruder. The green curve is for the nominal trajectory, and it can be seen that it violates the safety constraint. Moreover, it can be seen that both the classic and the scenario-tree MPC respect the minimum distance constraint. However, for the scenario-tree MPC, the distance is larger until around $51$ seconds. This is the time when the ownship passes behind the intruder, and practically, the scenario-tree MPC no longer affects the trajectory of the ownship after that. A detailed reason is shown in Figure~\ref{fig:MPC:3}, where the top-left (bottom-left) figure is at  $t = 35$s, and the top-right (bottom-right) figure is at $t = 57$s for the scenario-tree (classic) MPC approach. The corresponding predicted trajectories are shown in black. 

\begin{figure}[ht!]
\centering
\includegraphics[width=0.48\textwidth]{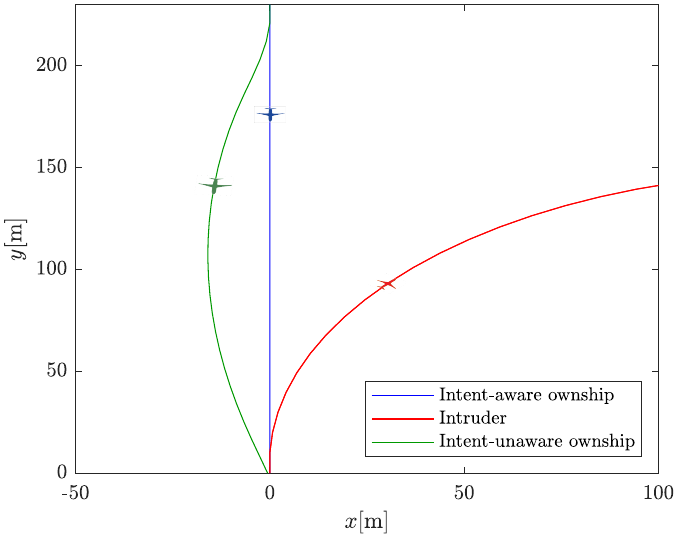}
\caption{Comparing the ownship path for the case of with (blue) and without (green) the intent information.}
\label{fig:MPC:01}
\end{figure}

Figure~\ref{fig:MPC:01} compares the ownship trajectory in scenarios with (blue) and without (green) the intent information. The intent information about the future trajectory of the intruder can be used to obtain a shorter path for the ownship while avoiding the collision. Note that in this figure, we have used a different encounter scenario than in the other simulation results to provide a clearer illustration of the importance and impact of the intent information.

%\begin{figure}[ht!]
%\centering
%\includegraphics[width=0.48\textwidth]{f6.pdf}
%\caption{Comparing the classic MPC trajectory (blue) and the scenario-tree MPC trajectory (black) for the ownship, the green area represents the destination for the ownship. The red trajectory corresponds to the intruder and follows Dubin's path.}
%\label{fig:MPC:1}
%\end{figure}

To demonstrate the robustness of the proposed method, we have conducted $20$ different simulations for the intruder, incorporating small bounded uniform uncertainties in the angular velocity as follows:
\begin{equation}\label{eq:stoch}
     u^2_k=D(\vect s^2_0,\vect s^2_T,k)+\omega, \quad \omega\sim \mathcal{U} [-0.5,\,0.5]\,\mathrm{deg}/\mathrm{s}, 
\end{equation}
where $\mathcal{U}$ is the uniform distribution accounting for environmental disturbances. Such variations can occur, for example, due to errors in the shared waypoints relative to the actual intruder path or deviations from the exact Dubins optimal path by the intruder. In the robust MPC scheme~\eqref{eq:RMPC}, the disturbance $\omega$ does not appear explicitly as a variable. Instead, uncertainty is modeled implicitly by constructing multiple branches in the scenario tree, each corresponding to a possible realization of the intruder’s angular velocity, including its nominal, minimum, and maximum values. This structure effectively captures a finite set of discrete outcomes, allowing the MPC scheme to anticipate different future trajectories and generate robust control actions accordingly.

\begin{figure}[ht!]
\centering
\includegraphics[width=0.48\textwidth]{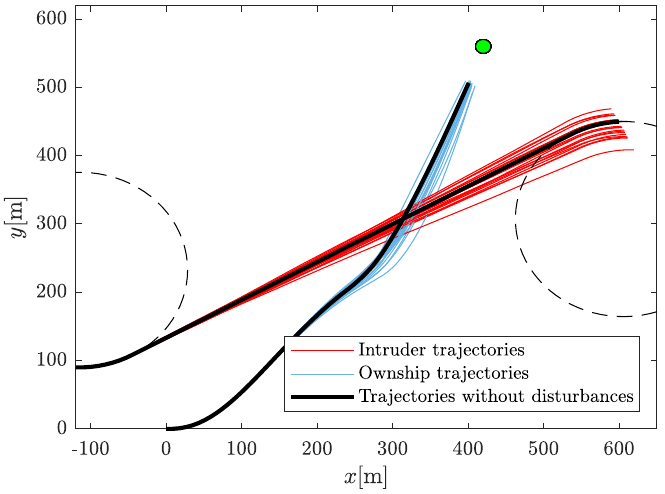}
\caption{The intruder (red) and the ownship (blue) trajectories for $20$ different cases arise from the intruder's angular velocity uncertainties. The nominal trajectories are shown in black.}
\label{fig:MPC:10}
\end{figure}

The resulting trajectories for the intruder (red) and the corresponding ownship paths (blue) are depicted in Figure~\ref{fig:MPC:10}. Although the disturbance at each time instance in \eqref{eq:stoch} is relatively small in magnitude, it leads to a fairly significant deviation in the terminal position of the intruder due to the propagation over time, with a difference of around $70$ meters as shown in Figure~\ref{fig:MPC:10}. Additionally, it's important to note that, for computational efficiency and to avoid overly conservative control inputs, the robustness of the scenario tree MPC is applied over a fairly short robust horizon rather than the entire maneuvering horizon.

\begin{figure}[ht!]
\centering
\includegraphics[width=0.48\textwidth]{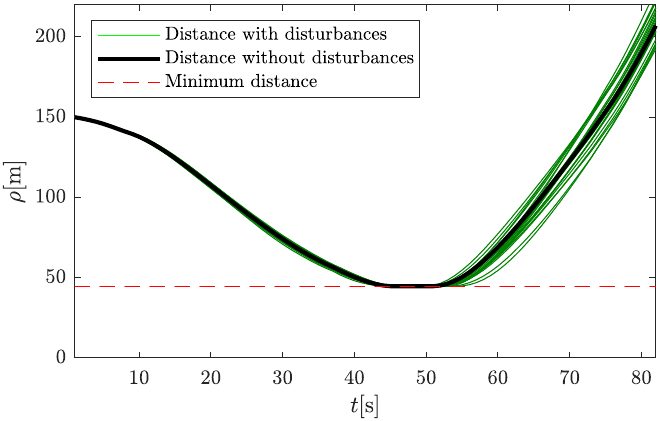}
\caption{The distance between the ownship and the intruder over time for different cases. The nominal distance is shown in black.}
\label{fig:MPC:11}
\end{figure}

Figure~\ref{fig:MPC:11} illustrates the evolution of the minimum separation distance between the ownship and the intruder over time for all simulated cases. The nominal distance profile (without uncertainty) is shown in black for reference. The figure shows that the proposed MPC framework respects the required safety distance constraint in all realizations under uncertainties.

\begin{figure}[ht!]
\centering
\includegraphics[width=0.48\textwidth]{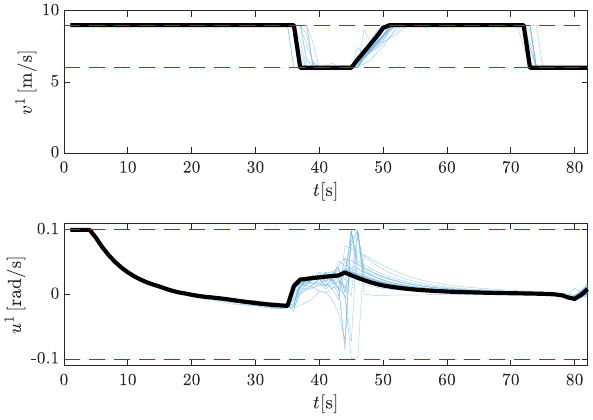}
\caption{The control inputs of the ownship (i.e., the linear velocity and angular velocity) and their boundaries for the scenario-tree MPC. The nominal velocities are shown in black.}
\label{fig:MPC:5}
\end{figure}

Finally, Figure~\ref{fig:MPC:5} shows the control input for the linear velocity and angular velocity and their boundaries for the scenario-tree MPC for all the uncertain cases. Note that in the CAS optimal path, the global optimal angular velocity is typically expected to be zero or at the boundary of the permitted interval. Figure~\ref{fig:MPC:5} generally aligns with this expectation. The slight deviation near zero may be attributed to the nature of the finite-horizon objective in the MPC scheme, which approximates the infinite-horizon objective, as well as to dynamic uncertainties.

Note that for the sake of clarity, in this paper, we have assumed that the ownship has priority, i.e., we apply the controller only to the ownship, while the intruder is assumed to take its optimal path, potentially with some uncertainties. There are works, e.g., in~\cite{stroeve2023matters}, where the authors consider CAS for both aircraft. In the MPC scheme, this can be achieved using distributed MPC, and it can be considered as a topic for future investigation.

\section{Conclusion}\label{sec:con}
This paper introduced a novel intent-aware collision avoidance system tailored for multi-agent aircraft engaged in horizontal maneuvering scenarios. By leveraging intent information in the form of waypoints or destinations, we proposed the use of scenario-tree model predictive control (MPC) to offer a computationally efficient and robust approach. Through simulations, these methodologies were compared to demonstrate their effectiveness in improving collision avoidance. In aircraft CAS, while dynamic programming (DP) is a common offline approach in this context, MPC can be performed entirely online during operation. This research highlights the significance of intent awareness in multi-agent systems and provides robust strategies for improving air traffic management, ensuring safe and efficient navigation in increasingly congested airspace. Future research will focus on employing DP and reinforcement learning to derive optimal policies and compare them with the method proposed in this paper. Additionally, more realistic scenarios, such as including response delays and uncertainty in the intruder dynamics can be considered in future research.

\bibliographystyle{IEEEtran}
\bibliography{IntentRMPC}
\end{document}